# Nanometer-Scale Deformations of Berea Sandstone under Moisture Content Variations

Eduard Ilin[1], Maxim Marchevsky[2], Irina Burkova[1], Michael Pak[1], and Alexey Bezryadin[1]

[1] Department of Physics, University of Illinois at Urbana-Champaign, Urbana, IL 61801

[2] Lawrence Berkeley National Laboratory, Berkley, CA 94720

*Sandstone mechanical stability is of key concern in projects involving injections of $CO_2$ in sandstone geological reservoirs, for the purpose of long-term storage. We developed a method to measure nanometer-scale deformations of sandstones in real time. We demonstrate that Berea sandstone, when hydrated, changes dimensions with a relative deformation of the order of $10^{-4}$. If the moisture content increases, sandstone samples exhibit an extension and if the moisture content decreases then the samples shrink. We also discover that, immediately after exposure to water, the sandstone temporarily shrinks, just for a few seconds, after which a slow extension begins, and continues until about half of the fluid evaporates. Such shrinkage followed by an extension has been observed also when the sample was exposed to acetone, Mineral Spirits or Vacuum Oil. The results are obtained using a high-resolution nano-positioner technique and, in independent experiments, confirmed using the technique of coda wave interferometry.*

## I. Introduction

It is known that exposure to water can increase the volume of the rock[1]. This effect differs significantly for different types of rock, with hydric dilation reaching values as high as 25% for certain claystones, while being substantially less pronounced for other types of stones such as granite or marble, with typical dilation values of 0.1% or less. There has been much interest in this process and a substantial amount of research activity, mostly devoted to potential degradation of mechanical properties of stones due to water dilation.

The question of the sandstone dimension change as a function of water content acquired a new increased significance recently due to the efforts to use sandstone geological reservoirs for long term storage of $CO_2$ gas[2]. Massive geological storage of $CO_2$ underground causes worrisome microseismic activities[3,4], which sometimes continue even when the stored gas leaves the injection site. It is known that the gas injection (as well as the extraction of a natural gas) leads to a massive relocation of water and brine underground, causing some areas to become dryer and some areas wetter. We propose that the observed seismic activity is due, at least in part, to water or brine relocations, casing the dimensional changes of the massive sandstone reservoirs and triggering avalanche-like events in critical tension states.

An overview of measured hydric dilatation for different types of rocks given in Ref.1 illustrates the universal nature of this phenomenon. At the same time, there is no commonly accepted interpretation of this property, thus further investigations are warranted, especially because the phenomenon is so important in geological projects as well as in any construction work involving stones. According to Ref.1, the largest relative extension values correlate with the greater amount of clay present in the stone, thus associating the pronounced dilation values with the swelling of clay minerals. For instance, hydric dilation measured for mudstone[5,6] reached up to 20 mm/m, while for high clay content sandstones dilatation values of up to 5 mm/m were reported[7,8] and the lowest reported value appears to be 0.1 mm/m[1]. On the other hand, rocks with very small clay content still exhibit some hydric dilation, although to a much lesser extent. Even clay-free rocks such as basalt or granite also expand when exposed to water, which suggest a universal mechanism for the dilation. It can also be suggested that not only the clay content, but the porosity correlates with the hydric extension coefficient. According to Ref.9, the larger observed expansion correlates with a greater percentage of micropores.

Significantly, the hydric dilation process is generally considered to be reversible, although the absence of measurable residual expansion after drying out does not prevent a possible loss of compressive strength, as was observed in some stones[10,11,12]. Here we report that the dimensions of the sample appear slightly reduced after complete drying of the sample. Thus, we present evidence that the cycle of wetting and drying is not completely reversible in terms of the sample dimensions and not only in terms of the mechanical strength. Such observed nanometer-scale changes of the dimensions can accumulate in geological settings and be a potential cause of the microseismic activity.

It was also demonstrated, in a series of experiments, that nanometer scale pores (nanopores) play a defining role in certain situations. For example, if the liquid condensation is considered, the capillary forces generated by nanopores are important[13,14,15,16].

To the best of our knowledge, no previous experiments attempted to measure the liquids - induced dimension-change effect down to nanometer scale precision of the Berea Sandstone, in real time. Such phenomenon of fast and short-time contraction upon saturation with liquids, which we report here, has not been reported previously. In this paper we fill this gap. Our experiments have been done on small samples, but we expect that same phenomena can occur and be even more pronounced on the geological scale.

## II. Experiment

In this experiment we investigate expansion and contraction of sandstone samples upon saturation with polar solvents water and acetone as well as non-polar solvents Mineral Spirits and Vacuum Oil.

The testing samples of the size of 25x8x3 mm were cut out of a large 100x100x20 mm sample and rinsed. This rock, quarried from Berea, Ohio, is a fine-grained sandstone composed mainly (~95% of solid volume) of sub-rounded to rounded quartz grains. Other constituent minerals include kaolinite (2%), microcline (1.5%), and muscovite (1.2%)[17,18,19]. The sandstone sample was mounted on the stable frame of a nano-positioner stage (Fig.1).

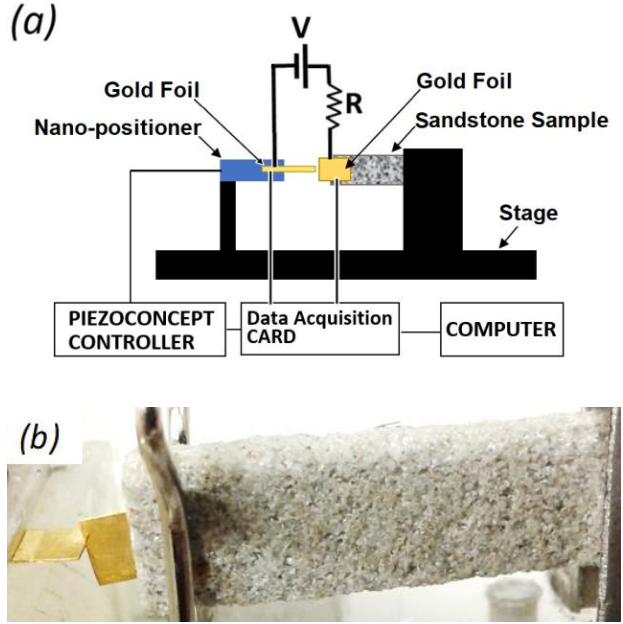

*Fig. 1. (a) Schematic of the experiment. The setup includes two gold foils forming so-called break junction, i.e., a contact with a mechanically controllable resistance. The left one is rigidly fixed to the moving part of the piezo-controlled nano-positioner stage of the apparatus while the right foil is rigidly fixed to a Berea sandstone sample. It can push the sample in the horizontal direction with the precision of a few nanometers. R is standard resistor of 1 MΩ. The voltage which needs to be applied to the* PIEZOCONCEPT *controller, in order to keep the contact resistance unchanged, is proportional to the extension of the sample. (b) A photograph of the setup. Two gold foils, in contact with each other, are visible on the left. One of the gold foils is attached to the sandstone sample, which is visible at the center.*

The sensitive part of the apparatus was a controllable break junction[20, 21] between two gold foils. One gold foil with the dimension 14 x 3.5 x 0.1 mm was attached to sandstone sample, while the second gold foil of similar dimensions was mounted on the moving part of the nanopositioner stage (Fig.1). The distance between the holder of the sample and the point where the gold foil was attached to the sample was ~13-18 mm. This distance was used as the effective length of the sample, when calculating the relative extension. The moving part of the stage was controlled by a piezo-positioner (PIEZOCONCEPT), which has a range of 30 microns for the total range of the control voltage of 10 V. The electrical resistance between the gold foils was monitored continuously during the experiment. A feedback mechanism was employed to maintain a constant electrical resistance of the break junction between the two gold foils. Using the piezo-positioner, the mechanism pushed the foils closer to each other if the electrical resistance between them increased and which pulled the gold foils apart from each other if the electrical resistance between them decreased with respect to a set value. If the feedback mechanism is activated, the magnitude of the shift of the gold foil always corresponds to the magnitude of the changing of the sample length, and the direction of the displacement of the gold foil is opposite the direction of the deformation of the sandstone sample. In other words, since the device was programmed to keep the contact resistance between the pair of the gold foils constant (Fig.1), therefore the shifts generated by the piezo-positioner were exactly equal to the changes in dimensions of the sandstone sample. The displacement of the piezo-positioner is proportional to the voltage applied to it. This voltage is actually measured to determine the sample displacement.

To reduce the effect of vibrations, the piezo-positioner was placed on a vibration isolation table (not shown). The contact resistance between the gold foils, i.e., the resistance of the break junction, was measured using a simple electronic circuit, which provided a fixed voltage of 0.03 V (using Tektronix PWS4205 voltage source) between the gold foils connected in series with a resistor and measured the voltage on the gold foils and on the resistor, thus allowing an accurate calculation of the contact resistance of the break junction. The resistor connected in series with a gold foils was 1 MΩ. To control the stability of the break junction a PIEZOCONCEPT device was used.

In our experiments, the feedback mechanism was activated and, after a delay of 5-10 min needed to verify that the system is stable in absence of external perturbations ~50 µl of liquid was added and the displacement of the piezo-positioner was recorded as a function of time.

The amount of added liquid was controlled by a micropipette and constituted 50 µl in all experiments. The liquid was always placed on the sample from above, generally taking on the order of several seconds for the liquid to soak the sample. There was no observed dependence of any measured properties on the sample positioning. In particular, it was experimentally determined that the soaking rate was not affected by placing the sample into a vessel containing the same amount of liquid as was added from the top in other experiments. Immediately preceding the experiments all samples were dried on the surface of an Isotemp heater for the duration of 2-3 hours at the temperature of $100^0$C. All measurements were protected from vibrational noise. To exclude the air convection effects, the sample was placed into a closed thermo-isolated chamber, which also contributed to maintaining stable temperature during the experiments. In all experiments the temperature was maintained at 20-$22^0$C and the relative humidity at 60%.

Our measurements have been done with polar solvents water and acetone as well as non-polar solvents Mineral Spirits and Vacuum Oil at room temperature, as well as with water heated to the temperature of 100°C (very near the boiling point).

The evaporation rate of liquids was measured using MXX-123 Denver Instrument electronic scale in a separate experiment. The precision of the electronic scale was ~1 mg. We first introduced 50 µl of liquid into the sample, immediately placed it on the scale and measured weight over time. The measurement shows, that as the liquid evaporates the weight drops and eventually returns to the original value when all liquid is evaporated.

## III. Results

**Hydric dilation of sandstones**

After adding the water at time $t = 10$ min counted from the start of the measurement we record changes of linear dimensions of the sandstone sample for ~150 minutes (Fig.2). When 50 µl of water was added the sample was completely soaked but no excess water was observed on its surface. The sample initially exhibited a rapid contraction, which continues for about two minutes and reaches an amplitude of the order of ~400 nm. After the initial and very rapid contraction, a relatively slow expansion begins, which continues up to the maximum extension of 600 nm (measured from the original size of the sample), achieved at time $t \approx 30$ min. After this, presumably due to the water evaporation, the sample starts to shrink again. As the sample reached its original size, the shrinkage of the sample did not stop but continued further, so at the end of the experiment the sample reached a somewhat smaller size than was measured before the hydration. The experiment was repeated three times, and the results are shown in Fig.2. In what follows we will show our data on the relative scale, as in Fig.3a. The maximum relative extension of this Berea sandstone exhibited a peak of about $4 \times 10^{-5}$, which occurred approximately 30 min after the initial soaking of the sample with water.

In a separate experiment, the mass of the sample was measured, continuously, after 50 μl of water was introduced into the sandstone using a micropipette. The time dependence of the measured weight of the fluid inside the sample is shown by the curves 4 and 5 in Fig.3b, which represent the dynamics of the evaporation process. As we compare the dilation curves 1, 2, and 3 and the weight curves 4 and 5, in Fig.3a and 3b, we notice that the contraction process appeared to continue longer than the drying process. The sample continues to shrink even after the sample weight returned to its original value, measured before the hydration. This observation indicates that some very small (<1 mg) amount of water may still be present even after the mass of the sample returns to its original value. Moreover, this small amount should be concentrated at the junctions between the grains, so that the effect of this remaining moisture on the dimensions is significant. In other words, our model uses the basic idea that the longer-time changes after all solvent evaporated is caused by some metastable concentration of solvent molecules adsorbed inside the sample, most probably inside the tiny pores present in the binding material, e.g., clay. Such view is reasonable since one cannot expect many pores being present inside the sand grains themselves. The observed approach of the additional mass to zero means that the accuracy of the weight measurement is not enough to account for the solvent molecules trapped between nanoparticles in the clay type binding material.

Below we will discuss a model which assumes that there is some cementing material (clay) present at the junctions between the sand grains. This clay has a relatively very small volume compared to all pores in the sandstone. But this cementing material is made of a matric of nanoparticles, and the nanopores in this matrix can trap water for a much longer time compared to the large pores ("macropores") present between sand grains in the sample. In such model the cementing material determines the size of the sample to a significant extent since it is located at the contact points (the junctions) between the sand grains. So, as this clay dries, the sample continues to shrink.

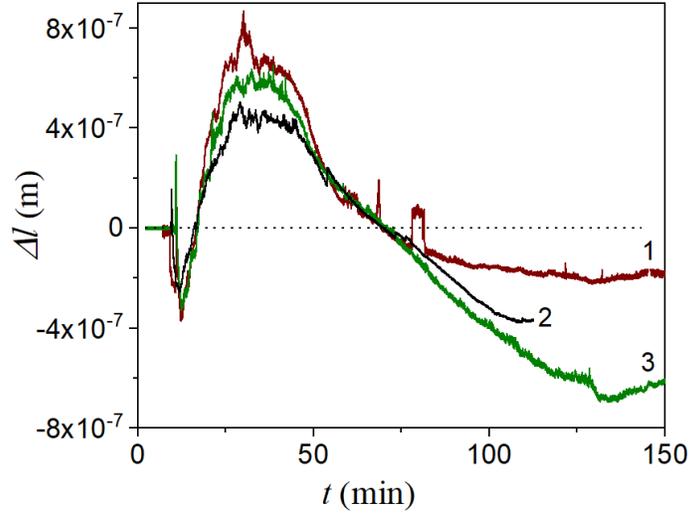

*Fig. 2. The change of the length of the sandstone sample versus time. The sample was saturated with water at time t=10 min. The amount of water was 50 µl. The measurement was repeated three times; thus three curves are shown. Right before each measurement the sample was completely dried on a hot plate (~100°C).*

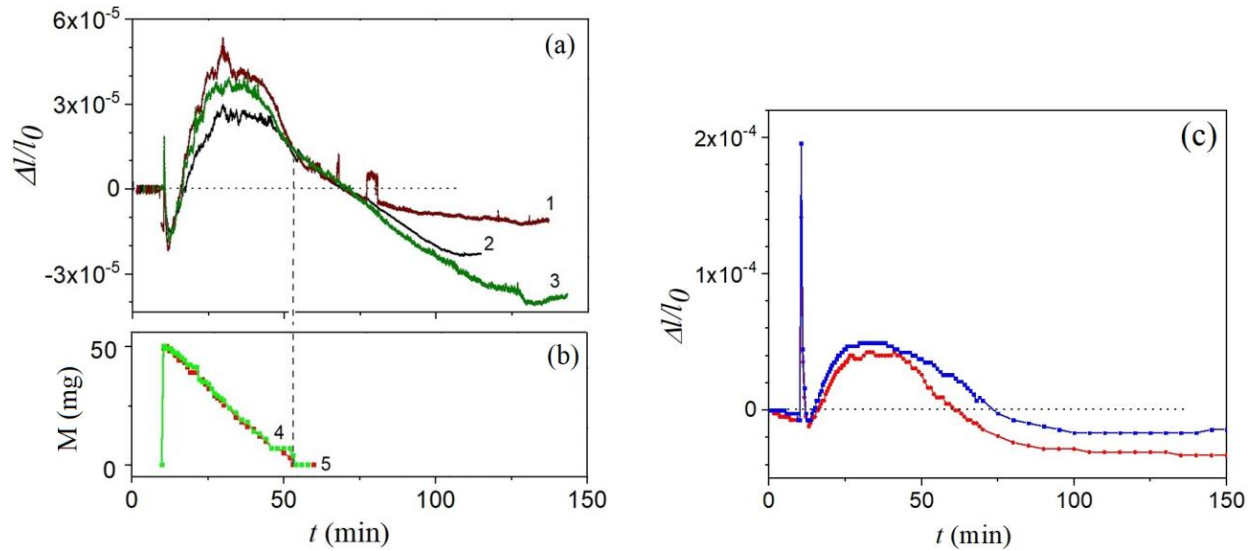

*Fig. 3. (a) Relative change of linear dimension of the sandstone sample saturated with water at room temperature plotted versus time. The amount of water was 50 µl. The horizontal time axis in (a) and (b) is the same. The three curves, 1, 2, and 3, represent three similar independent measurements. (b) The curves 4 and 5 represent the measured weight of the water, i.e., the differential weight, inside the sample versus time. To measure these curves, 50 µl of water was introduced into the sample at t=10 min and the sandstone was left on the scale to dry, as its weight was measured periodically. (c) Relative change of the sample length after it was saturated with water having the temperature of ~100°C.*

**Effects of temperature on hydric dilation**

In a separate experiment we tested the effect of temperature changes. For this purpose we first heat water to near its boiling point (100°C) and then placed 50 µl of this hot water on the sample to soak it. The sample expanded sharply immediately after the heated water was added, the maximum relative extension observed was ~$1.9 \times 10^{-4}$ (Fig.3c). After about 2 or 3 minutes the sample returned to its original dimensions. The subsequent behavior of the sandstone sample was similar to the hydration tests done with water at room temperature. The sample expanded for the first 30 minutes, followed by contraction to initial dimensions in around 50 – 60 minutes (Fig.3c). In this test, the initial behavior of the sample is likely explained by the thermal expansion due to the heat transfer from added water to the sample. Our qualitative explanation (to be discussed later, in more detail) is that the introduced heated water comes into contact with the surface of the grains first, and, what is of key importance, with the clay present at the junctions between sand grains. The clay temperature goes up and it expands quickly. After a few minutes the heat becomes distributed evenly and much of it goes into grains. But since the volume of the grains is much larger than the volume of the clay, the resulting equilibrium temperature deviates only by a small amount from the room temperature. Under "small" we mean that the difference is much smaller than the average value of the absolute temperature.

We estimated linear dimensional changes of the sample using the following expression: $\Delta L = \alpha_L L \Delta T$, where $\alpha_L$ is the coefficient of linear thermal expansion and $L$ is the sample length. Based on the literature [22] we assume $\alpha_L = (10.1 \div 12.1) \times 10^{-6}$ K$^{-1}$. The change of the sandstone sample temperature was estimated using the heat balance equation. The heat capacity of sandstone is $0.92 \times 10^3$ J/(kg K) [23]. Our estimate shows that the temperature of the sample has to increase by approximately 14 K after 50 mg of hot water (100°C) is introduced into it. Taken into account this change of the sample temperature we calculate the change of the linear dimension of the sample: $\Delta l/l_0 = (1.4 \div 1.7) \times 10^{-4}$ whereas experiment shows $\Delta l/l_0 = (1.4 \div 1.9) \times 10^{-4}$ for 100°C water. Thus, the estimate is in good agreement with the measurements.

**Acetone dilation of sandstone**

Another question of interest is whether the sandstone extension, observed upon water saturation, is uniquely connected to water. Thus, similar tests have been performed with acetone (Fig.4). When acetone was introduced to the sandstone sample, the sample initially exhibited a sharp contraction, followed by a slower expansion stage, which continued for roughly two minutes until reaching the maximum size, after which the sample contracted for 60 – 70 minutes. Exposure of the sandstone to room temperature water resulted in a similar behavior. But the processes of expansion is faster with acetone than with water. As with water, after the acetone is evaporated, the sample continues to contract. The time dependence of the weight of the acetone inside the sample is shown by the curves 4 and 5 in Fig.4b, which represent the dynamics of the evaporation process.

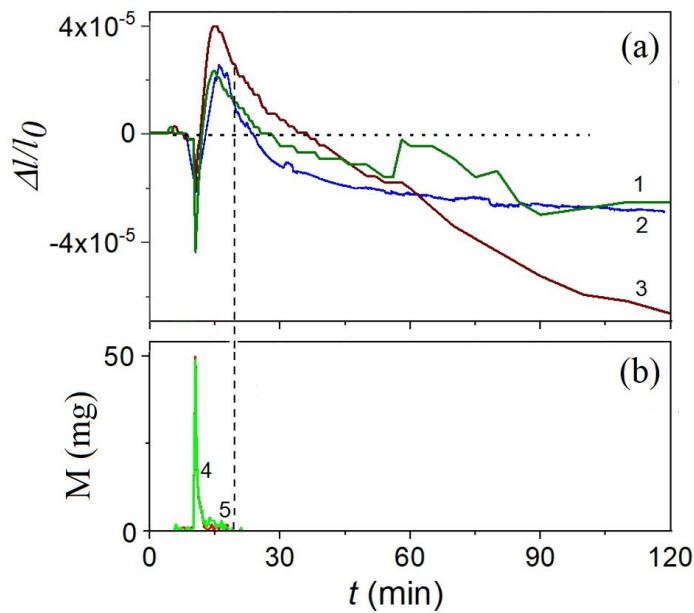

*Fig. 4. (a) Relative change of linear dimension of the sandstone sample, which was saturated with acetone at room temperature, plotted versus time. The three curves, 1, 2, and 3, represent three similar independent measurements. (b) Evaporation of acetone: The time dependence of the measured weight of the acetone inside the sample. The initial amount of acetone was 50 μl. The horizontal time axis is the same both for (a) and (b).*

As with water, the contraction phase continues long after the acetone is evaporated and the weight of the sample returned to its original value, with the precision of our scale. As with water, our explanation is that a minute amount of acetone remains trapped in the nanopores of the cementing material concentrated at the junctions between the grains and holding them together.

**Dilation of sandstone with non-polar liquids**

To elucidate the effects of liquid polarity on sandstone dilation, additional experiments were performed with non-polar solvents Mineral Spirits and Vacuum Oil.

Similar to the case of polar solvents, the addition of Mineral Spirits leads to a short-term contraction of the sample, followed by much slower expansion (Fig.5a). The expansion stage up to the maximum size lasted for approximately 7 minutes, followed by contraction of the sample. This behavior is similar to what was observed in water and acetone experiments, although the expansion process with Mineral Spirits occurred faster than with water, but slower than with acetone (Table 1). Just like in the case of polar liquids, after the complete evaporation of the solvent the contraction of the sample continued.

Table 1. Liquid evaporation times and approximate durations of the observed expansions.

|  | Water | Acetone | Mineral Spirits | Vacuum Oil |
|---|---|---|---|---|
| Evaporation time, min | 41-45 | 9-9.5 | 16-16.5 | >120 |
| Time to maximal expansion, min | 20-25 | 4.7-5.9 | 6.7-7.8 | >120 |
| Expansion time, min | 52-57 | 13.8-24.5 | 24-28 | >120 |

Another non-polar liquid used was Vacuum Oil. On introduction of the liquid, an initial short-term contraction (approximately one minute in duration) was followed by at first abrupt and then considerably slower expansion (Fig.5a). The expansion continued for the entire duration of the experiment. Weighing the sample after the addition of 50 μl of the liquid demonstrated no changes the mass of the sample in 120 min (Fig.5b).

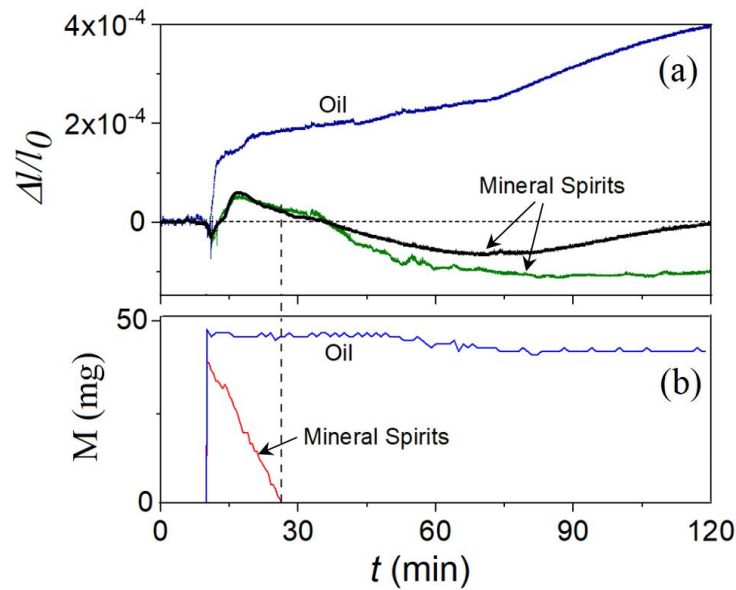

*Fig. 5. (a) Relative length change of the sandstone sample plotted versus time. The sample was saturated at t=10 min with either mineral spirits or a vacuum machine oil at room temperature. The time axis for (a) and (b) is the same. (b) The time dependence of the weight of the either mineral spirits or a vacuum machine oil inside the sample, determined as the current sample weight minus the weight of the sample measured before the introduction of the fluid. The amount of liquid used was 50 μl.*

**Acoustic measurements**

We also conducted a series of experiments using an acoustic technique of coda wave interferometry (CWI)[24,25] to confirm the character of dimensional changes in the sandstone sample subjected to a moisturizing process. CWI is a technique based on ultrasound propagation, which is widely used for detection of small local changes within an inhomogeneous medium[26]. If a sound wave induced as a short pulse passes through a complexly shaped sample with multiple internal scatters and interfaces, it travels along multiple complex trajectories exhibiting multiple reflections along its propagation path from sample internal inhomogeneities and edges. Before being fully absorbed, the wave travels along the sample multiple times, effectively probing its interior. Such decaying waves are called coda waves (Fig.6a). There are highly repeatable, even if the medium is very disordered, provided that the sample does not change with time. This means that by applying the same pulse to the sample one gets the same time dependence of the resulting vibrations. Yet, if something changes within the sample then coda waves exhibit a strong variation. Thus, the technique is a sensitive probe for small internal changes in the sample. For example, if the sample elongates, it takes longer for the wave to pass through the sample and the pattern shifts in time towards larger times.

To test the effect of moisture on the sandstone, the sample was mechanically excited with a 2.4 µs-long acoustic pulses applied with a piezoelectric transducer glued to the sandstone surface, and the resulting reverberating acoustic waveform was acquired by a second piezo-transducer at the opposite end of the sample, and recorded using a PicoScope6000 Digital USB oscilloscope. We monitored variations of the received waveform as a function of time by pulsing the sandstone sample 10 times a second and calculating the time shift corresponding to the maximum cross-correlation between the initial (reference) and subsequent waveforms. The initial waveform was recorded when the sample was dry and equilibrated at room temperature (Fig.6a).

In the first test, a hot soldering iron was brought to the ~2 cm distance from the sandstone sample and held in place, causing its surface temperature to rise by ~1 K over the 20 s time interval. The observed time shift amplitude was $-2.5 \times 10^{-7}$ s (where negative sign corresponds to a delay of the reverberating wave relative to the reference one). This is due to an increase of the time needed for the sound waves to traverse the sample.

In the next test, moisture-saturated air at 36°C was blown on the sandstone surface via a straw, while the ambient air humidity was at ~40%. In this experiment, the time shift was $-1.2 \times 10^{-6}$ over 5 s, while no noticeable surface temperature change was observed. Thus, the moisture-induced effect is nearly an order of magnitude more significant than the thermal one. Notably, in both experiments the time shift was positive along the time axis, corresponding to an overall increase of acoustic wave travel time. In the first experiment, we associate this increase with the thermal softening of sandstone Young's modulus that reduces sound velocity, while in the second experiment the moisture-driven volume expansion of the sample is causing accumulation of time delay in the reverberating wave. In the first experiment the time shift was fully reversible and returned to a near-zero value when the sample returned to uniform ambient temperature. In the second test with the high humidity air, the time shift decreased in amplitude as the sample was drying and approached zero. Then the time shift changed the sign and saturated at a slightly positive value suggesting that the sample developed a residual shrinkage, similar to the experiments involving nanometer scale direct measurements, discussed above.

This effect was confirmed in the third experiment, where a drop of acetone was deposited on the sample. The result (Fig.6b) reproduces the general dimensional change dependence shown in Fig.2 of the paper quite well: expansion is followed by contraction and shrinkage relative to initial dimensions. The only notable difference is the absence of the initial contraction "spike" upon introduction of the liquid. It appears that the sample stayed contracted indefinitely, and never returns to the original state.

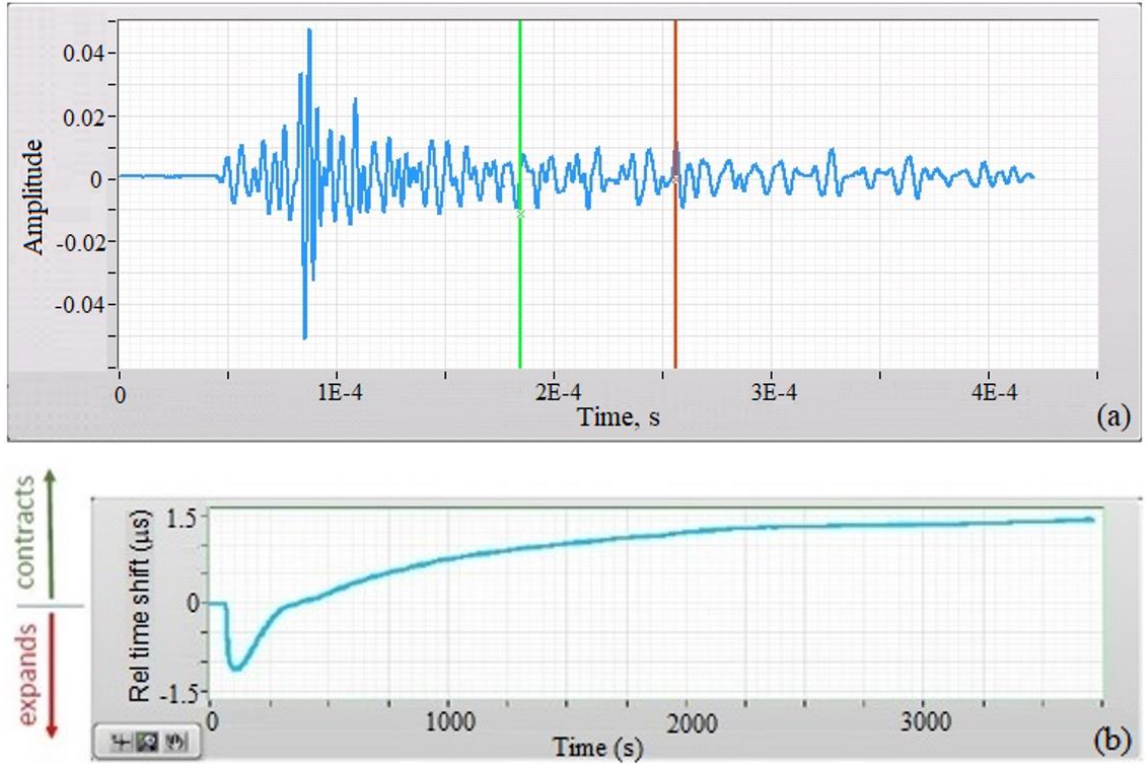

*Fig. 6. Acoustic measurements of coda in acetone-soaked sandstone. (a) A snapshot that shows the actual coda waveform monitored in the experiment, as well as its selected portion, which is used to compute the plot in (b). The selected portion is roughly 20 reverberation periods from the beginning of the coda, and its length is ~ 10 periods. But as it is not exactly periodic, these numbers should be taken as a rough estimate. (b) The time shift of the reverberating sound wave as compared to the reference wave measured on the dry sample. As acetone is deposited onto the sample, the time shift becomes negative. This is consistent with the expected increase of the sample dimensions. As the sample dries out the time shift increases and eventually becomes positive, indicating sample contraction to a somewhat smaller dimensions compared to the initial ones.*

## IV. Discussion

Our newly developed experimental techniques appear to reliably measure dimensional changes with a relative deformation of the order of $10^{-4}$, and with an absolute resolution of ~10 nm or better, depending on the levels of vibrations, voltage control and temperature stability. The measured expansion is generally consistent with previously observed results for sandstones[1]. However, there are several new findings.

One such observation is the aftereffect, which is the shrinkage of the sample relative to its original size after complete desiccation. Additionally, there is a notable mismatch between the timing of desiccation observed by monitoring the weight of the sample vs. changes of the sample size. The dimensional changes appear to continue considerably longer in time after the drying process, as measured through the weight change, appears to be completed. We propose the following simple explanation for this observation. According to Ghurcher P.L. *et al*[27], Berea sandstone composition is approximately 85-95% of quartz, 6-8% clay and 1-2% dolomite. The sandstone we use contains 2% of kaolinite[17,18,19]. The bulk of pore space in sandstone consists of open space between the individual grains of sand forming the sandstone (Fig. 7, 8). These pores are relatively large since the grain size is of the order of 100 µm (Fig. 8). This is consistent with previously published data[28]. These macro-pores are well interconnected and are easily and quickly filled with water immediately (on the order of a few seconds in our case, because of the mm sample size in our experiment) after soaking, while water is also removed from this pore space relatively quickly when the sample is dried (over a time of ~53 min according to Fig. 3a and 3b). Due to the large size of these pores (Fig.7, 8), the presence of water there contributes very little to the hydric expansion of the sample. Another type of pores is present in the cement material (clay) gluing the sand grains together[29,30] (marked "Clay" in Fig.7).

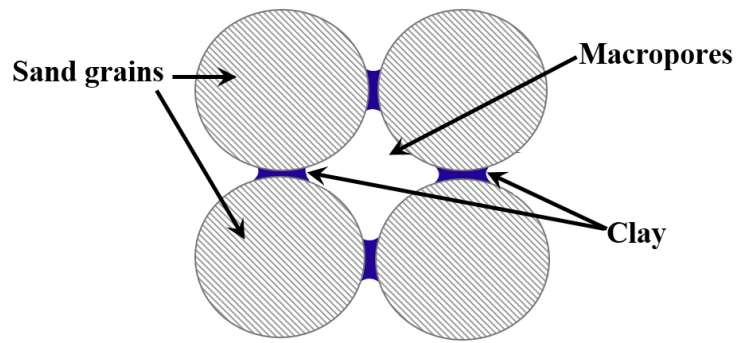

*Fig. 7. Schematic representation of sandstone microscopic structure. Typical dimensions of sand grains and the macropores between them is of the order of 100 µm. The "nanopores" are present inside clay (blue); they are not shown explicitly in the drawing. In our sandstone, the volume of the clay (blue) is much smaller than the volume of the macropores (white) present between sand grains.*

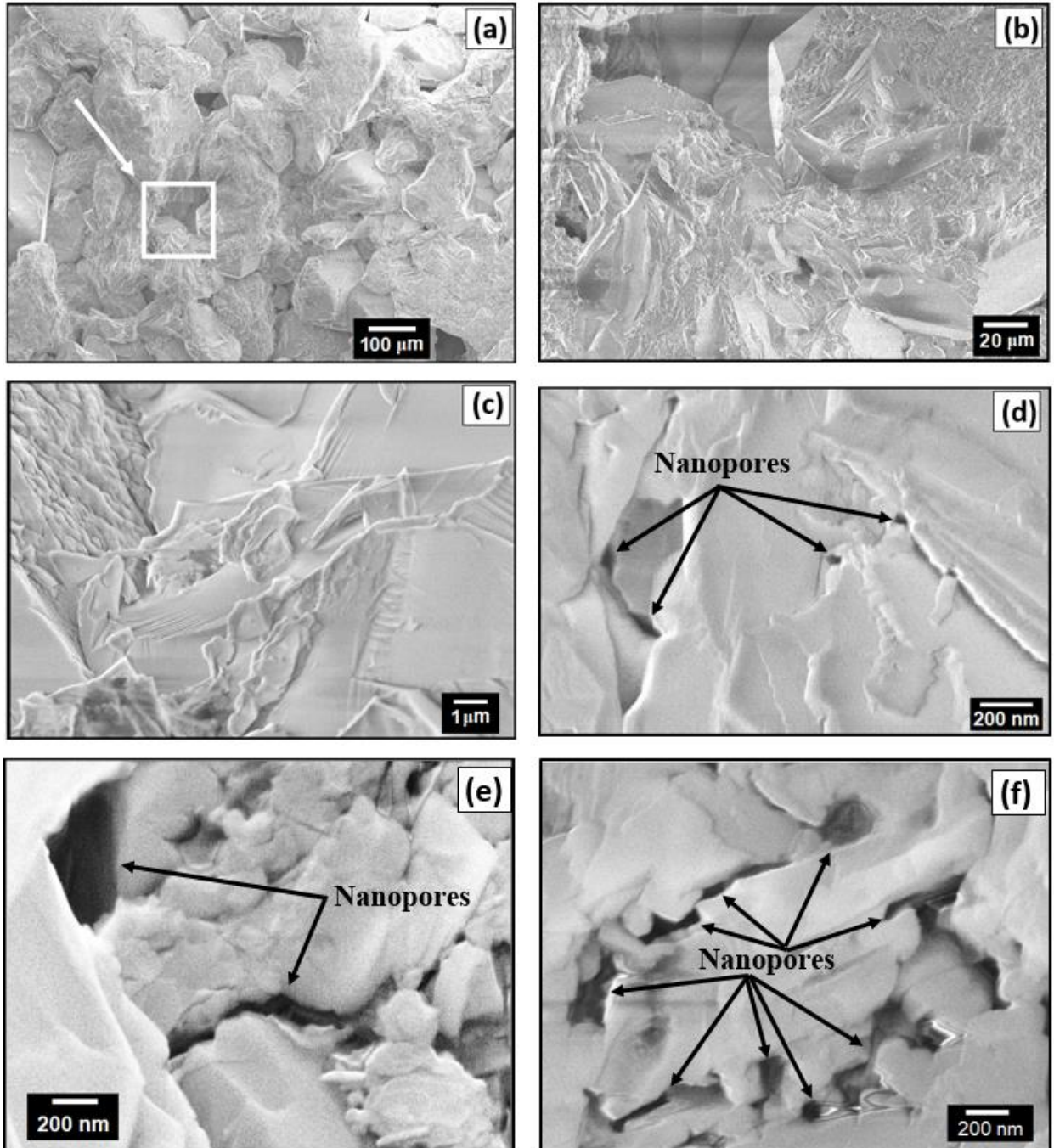

*Figure 8. Example micrographs of the sandstone surface obtained by a scanning electron microscope with different magnification. Pictures b, c, d, e, f are magnifications of the area enclosed in a square on Pic. a. Nanopores are marked by arrows on Pic. d, e, f. The highest magnification image (Pic. e) shows very pronounced nanopores, which appear as black areas on the image. Another example of nanopores is given in (Pic. f); they are indicated by arrows.*

In Fig.7 the volume of the clay (blue) is illustrated as being much smaller than the volume of the macropores. It can be seen from the micrograph on Figure 8 that the micropores between individual sand grains are much larger in size than the cementing elements of clay. It was estimated[17] to be only 2% of the volume of the macropores (shown as white-color regions in Fig.7). This clay has some pores also, of with a much smaller size[30]. They will be called "nanopores" since their dimensions, defined by the dimensions of the typical clay particles, is in the nanometer range[30,31](Fig. 8d,e,f). While nanopores occupy a much smaller volume, it is the penetration of water into these pores that primarily causes the hydric expansion of sandstone, because clay serves as a sort of "glue" connecting the sand gains[29]. Thus, if the clay expands the sample expands by the same amount. The mechanism of the clay expansion is related to the small size of clay-forming particles, which is of the order of a micrometer or less. Because of this small size water molecules adhere to surface of clay particles and push them apart. Even a single layer of water molecules can produce a significant expansion. The size of the water molecule is ~0.3 nm. Thus, if the particles of clay are 100 nm in size[30,31] the relative expansion of the clay interlayers would be 0.003. Since the size of clay regions may be about 3% of the size of the sand grains, the rough estimate for the sample relative extension in this case is 0.0001, which matches the results provided above. Thus, our suggestion is to explain the observed extension by the penetration of a small amount of water between the clay particles (or clay layers), on the amount of possibly just one or a few layers of water molecules. Such a small amount appears to produce noticeable results since the clay particles are very small. Each particle of clay is surrounded by water like a cushion, thus pulling clay particles apart. Of course, this mechanism is universal and would apply to any liquid which wets the surface of clay particles. Thus, it works with acetone, Mineral Spirits, and Vacuum oil, for example, as we have reported above.

In particular, the structure of kaolinite, which might be present between the grain[17,18,19], is composed of silicate sheets ($Si_2O_5$) bonded to gibbsite layers (aluminum oxide/hydroxide layers ($Al_2(OH)_4$) [32,33]. The silica sheet and the gibbsite sheet are held together by hydrogen bonding between the oxygen atoms at the tips of the silica sheet and the hydroxyl ions of the alumina sheet, forming the common interface between the two sheets. Hydration effects immediate surroundings of these ions, leading to change of inter-sheet physical dimensions, because the hydrogen bonding is affected[34].

On the other hand, the grains are made of non-porous quartz, so they do not expand by themselves. It also takes substantially more time to rid these small clay nano-pores of water in the drying process[35]. This is consistent with the observed time dependences of hydric expansion and hydration (by mass) processes shown in Fig. 3a and 3b. Most of added liquid quickly penetrates the large pores but it has no immediate effect on the sample size. A small amount of liquid slowly begins to penetrate the clay nanopores, causing the expansion, while much large amounts of liquid are already leaving the sample in the drying process[36]. After around 25 min water penetrates the nanopores in the clay and the sample shows the maximum extension. At this time, as sample weight monitoring shows (Fig. 3a and 3b), about 50% of water filling the large macro-pores is evaporated, and some empty space appears in the pores of the sample. Thus, the water trapped in the clay nanopores begins to evaporate into partially empty macro-pores between the sand grains[37] and so the sample starts to shrink. It takes much longer for the water to leave the clay nanopores[38], so the sample exhibits overall dimension decrease over a long time, of the order of 100 min. Apparently such flashing of the nanopores in the clay cementing material causes some removal of its constituents so a small after-effect is observed, which means that after complete drying the size of the sample is slightly smaller than the original size, measured before the hydration.

The initial very rapid change of the dimensions of the rock as is injected into it is more quite difficult to explain. We speculate that this effect might be temperature-related, since its duration coincides with the duration of the hot-water initial expansion effect, which is shown in Fig.3c. Compare, for example, the initial expansion peak in Fig.3a and the peak observed in the experiments where the sample was saturated with hot, ~100°C, water as shown in Fig.3c. The duration of the initial dimensional change is of the same order of magnitude. Thus, a thermal initial effect appears plausible. Another possible explanation is that the capillary forces are responsible for the initial contraction, as the fluid penetrates the sample. We think that this explanation is less probable since the water penetrates the sample much faster, as can be observed visually since the sample changes the color when wet.

## V. Conclusions

We have employed a nano-positioner in combination with electrical break junction to measure minute changes in the linear dimensions of sandstones at the nanometer scale in real time. This sensitive technique was applied to study small changes of the dimensions of sandstones as they were saturated with polar solvents water and acetone, as well as non-polar Mineral Spirits, and Vacuum oil. We find that, after Berea sandstone is soaked with water, its length increases over a time scale of ~25 min. The extension was not irreversible. On the contrary, as the water (or acetone and Mineral Spirits) evaporated from the sample, its length decreased again, reaching a value slightly smaller than the original size. Two unexpected effects have been discovered. The first one is the apparent continued contraction of the samples even after all water was evaporated. The time span of the water evaporation was determined by weighing the sample. The weight returned to its original value after approximately 52 min (in experiments with water). Surprisingly, the reduction of the sample dimensions continued further, and as the drying of the sample continued, its dimensions reduced below the original value.

To explain this observation, we propose a model that assumes that the clay cementing the sand grains together into a rigid matrix contains some extra small pores, the nanopores. It takes a much longer time to fill these clay nanopores with water, namely about 25 min in our samples. By our estimate, these nanopores are so small that even a single layer of water molecules penetrating into them would push clay nanoparticles apart and contribute enough extension to the entire sandstone sample that it would match our experimentally observed extensions. Also, when the bulk of water has evaporated and the weight of the sample appears to return, within experimental precision, to the initial weight, some water might still be present in the clay nanopores. Eventually all of it evaporates, but this takes a significantly longer time. The water flashing possibly changes the internal structure/composition of the clay nanopores, thus causing the observed small reduction of the sample dimension after each wetting-drying cycle.

**Acknowledgments**

This work was supported as part of the Center for Geologic Storage of $CO_2$, an Energy Frontier Research Center funded by the U.S. Department of Energy, Office of Science, Basic Energy Sciences under award number DE-SC0C12504.

The authors would like to thank S. Frailey and J.S. Popovics for useful discussions.